%% file: paper.tex
\documentclass[preprint2]{aastex}
\slugcomment{Astrophysical Journal, October 10 issue}
\lefthead{El-Zant, \& Shlosman}
\righthead{}  
\input epsfig

\def \m3{{\rm Mark III}}

\begin{document}

\title{LONG-LIVED DOUBLE-BARRED GALAXIES:\\ CRITICAL MASS AND LENGTH SCALES}
\author{Amr A. El-Zant}
\affil{Theoretical Astrophysics 130-33, California Institute of  Technology, Pasadena, CA 91125
\\email {\tt elzant@tapir.caltech.edu}}
\and
\author{Isaac Shlosman}
\affil{Department of Physics \& Astronomy, University of Kentucky, Lexington,
   KY 40506-0055\\ email: {\tt shlosman@pa.uky.edu}}

\begin{abstract}
A substantial fraction of disk galaxies is double-barred. We analyze the dynamical 
stability of such nested bar systems by means of Liapunov exponents, by fixing a
generic model and varying the inner (secondary) bar mass. We show that there exists 
a critical mass below which the secondary bar cannot sustain its own orbital structure, 
and above which it progressively destroys the outer (primary) bar-supporting orbits.
In this critical state, a large fraction of the trajectories (regular and chaotic) are 
aligned with either bar, suggesting the plausibility of {\it long-lived} dynamical states
when secondary-to-primary bar mass ratio is of the order of a few percent. 
Qualitatively similar results are obtained by varying the size of the secondary bar,
within certain limits, while keeping its mass constant. In both cases, an important role appears 
to be played by chaotic trajectories which are trapped around (especially) the primary 
bar for long periods of time. 
\end{abstract}


\section{Introduction}

If long-lived double-barred galactic systems were not observed, explaining the fact 
would hardly pose a major theoretical challenge. It would be sufficient 
to suppose a situation whereby generic trajectories transit erratically between 
motion characteristic of resonant, bar-supporting  orbits of the inner and outer 
bars  --- their shapes, in the process, supporting neither bar.  Indeed, in general,
trajectories that can switch between qualitatively different modes of motion, such as
those of a pendulum near the vertical point or a ball near the top of the hill of a 
double-well potential, will be chaotic, unless the system  exhibits exceptional 
symmetry.\footnote{In the case of the pendulum, it suffices that it be unperturbed.} In 
fact the resulting ``homoclinic'' phenomena (which guarantee at least transiently erratic 
motion) can be taken as to 
define the phenomenon of dynamical ``chaos,'' in both its conservative and Hamiltonian 
manifestations  (Holmes 1990; Ruelle 1989). In a galaxy with two bars tumbling with 
different pattern speeds, there are, in general, no special symmetries, and no 
corresponding smooth integrals of motion in the phase space.
 
Nevertheless, there is reason to believe  that  structures, such as nested bars, can 
materialize. Shlosman, Frank \& Begelman (1989) have analyzed possible formation mechanisms.  
An argument was put forward that nuclear bars are {\it secondary} dynamical features which form 
due to gravitational instabilities in the gas accumulation within the central kpc and 
subsequently affect the background stellar component. Probably the most dramatic result of this 
analysis was that double bar systems spend a large fraction of their lifetime in a dynamically 
decoupled state, characterized by substantailly different pattern speeds of each bar.  

High-resolution ground-based observations have revealed a number of galaxies with a sub-kpc 
secondary stellar bars (e.g., Buta \& Croker 1993; Shaw et al. 1995; Friedli et al. 1996; 
Jungwiert, Combes \& Axon 1997; Mulchaey \& Regan 1997; Jogee, Kenney \& Smith 1998; Erwin 
\& Sparke 1999; Knapen, Shlosman \& Peletier 2000; Emsellem et al. 2001). The first statistics 
on nested bar galaxies, has been limited exclusively 
to stellar bars due to superior resolution in detecting the stellar light distribution and 
kinematics. The most comprehensive, so far, {\it HST} survey of 112 galaxies finds that
in excess of 20\%---25\% of disks host double bars, and about 1/3 of all barred galaxies
host another (nuclear) bar  (Laine et al. 2002). The former can even reach 40\% 
(Erwin \& Sparke). A clear indication that nested bars indeed tumble with different 
pattern speeds comes from their random mutual orientation (Friedli et al.) and, indirectly, 
from the bimodal length distribution of bars in these systems (Laine et al.)

The frequency of detection of double-barred systems suggests that, at least some of them, 
can be relatively long-lived. Since we do not expect that they can be built of trajectories 
that can transit between the modes of motion of the two subsystems, i.e., of untrapped chaotic 
orbits, we conjecture that such systems are composed of orbits that are trapped either around 
the outer (primary) or the inner (secondary) bar. Under this assumption (verified {\it a 
posteriori}), the system is made of regular orbits confined to each bar and trapped chaotic 
orbits in the vicinity of the regular regions.

Within the context of the KAM theorem (e.g., Arnold 1987), the continued 
stability of quasiperiodic solutions in  a dynamical system
requires that external perturbations be sufficiently small, in which case most of these 
orbits remain, even if slightly deformed. Our definition of decoupled bars as simply tumbling
with different pattern speeds does not necessarily imply this. This is because the gravitational
quadrupole interaction between the bars can, in principle, be strong enough 
so as to destroy a large fraction of the regular trajectories --- which, in general,
can result in the dissolution of either, or both, bars. Thus one has to ask under what conditions the 
supporting trajectories of each bar are also KAM-stable under the influence of the second bar 
perturbation. This issue is addressed here. 

\section{Model and Method}

To answer the questions posed above, we define a generic model and vary its parameters. 
We find it appropriate to examine motion in a fixed potential, where bar parameters
can be set at will, rather than 
attempt $N$-body experiments, and choose variants of one of the models for a double-barred 
galaxy previously investigated by Shlosman \& Heller (2002). These consist of halo and bulge 
modeled as Plummer spheres (the former with a large core of 10~kpc) and a Miyamoto-Nagai (1975) 
disk, supplemented by two Ferrers (1877) bars of order 1, with the secondary bar rotating 
$8.3$ times faster than the primary one. The secondary bar mass has been varied
by factors 2 and 4, above and below its mass in the Model~1 (hereafter generic 
model) of Shlosman \& Heller. Qualitatively similar results are obtained if one varies the surface 
density of the secondary by changing its size around generic value, as long as the relevant 
primary bar orbits are still affected by the quadrupole moment of the secondary bar.
The generic ratio of secondary-to-primary bar masses is 0.047 and their surface density ratio about 8. 
The bar size ratio is 0.08, which positions the secondary corotation between the inner Lindblad 
resonances (ILRs) of the primary. Each bar comprises about 20\% of the total mass within bar radii.

We employ Liapunov exponents (see El-Zant \& Shlosman [2002] for complete details) in order to 
examine the stability of trajectories, and confine ourselves, in this first exploration, to 
two-dimensional motion. We choose a mesh of a hundred initial positions, equally spaced 
along the primary bar major axis. Each position is a starting point for a hundred trajectories with 
normal velocities equally subdivided in the range between $1.25$ times the local rotation velocity 
(excluding the bars' contributions) to a hundredth of this value. Such initial conditions 
should adequately describe bar-supporting orbits, i.e., those aligned with each bar. These will mostly
be parented by generalizations of the so-called closed $x_1$ orbits of single bar systems.  
As such, their symmetry requires that they, at some stage, cross the  major axis of the bar with 
normal velocities. Starting 
from these initial values, we advance the trajectories for $50,000$ Myr. The rationalization for 
this particular value was discussed in detail in El-Zant \& Shlosman.

\section{Results}

The rationale behind the present work is straightforward: given a primary 
bar which exists as a long-lived configuration, we are interested in investigating 
the range of parameters (if any) for which secondary bars are sustainable, 
yet do not sufficiently interfere with the primary bar dynamics, so as to destroy it. 

Our findings are summarized in Fig.~1, where we display the five models with increasing 
secondary bar mass (by a factor of 2 each from top to bottom). The middle panels (third
from the top) are those 
of the generic model. Specifically, (1) grayshades in the left column show values of the
Liapunov exponents, our ``measure of chaos;'' (2) the middle column marks trajectories 
whose maximal extension along a bar is twice or more their extension normal to a bar, for both 
the outer ({\it plus signs}) and inner ({\it dots}) bars; and (3) the right column
exhibits greyshades of axial ratios of orbits. Liapunov timescales are varied from $10^4$~Myr 
(white shades), corresponding to regular orbits, to $10^2$~Myr (black shades) corresponding 
to highly chaotic orbits. The axial ratio, $p\equiv a/b$, has values greater than unity, 
for all bar-supporting orbits, in  the appropriate bar frame. 
However, it will be very close to unity in a given bar frame, {\it if the trajectory is not 
trapped by that bar}. It is appropriate, therefore, to take the maximal value 
of $p$ in each of the frames. In scaling the greyshades the following  limits
are employed: white corresponds to $p \ge 3$  and  black to $p \le 1$. 

\begin{figure*}[ht!!!!!!!!!!!!!!!!!!!!!!!!!]
\vbox to7.0in{\rule{0pt}{4.8in}}
\includegraphics{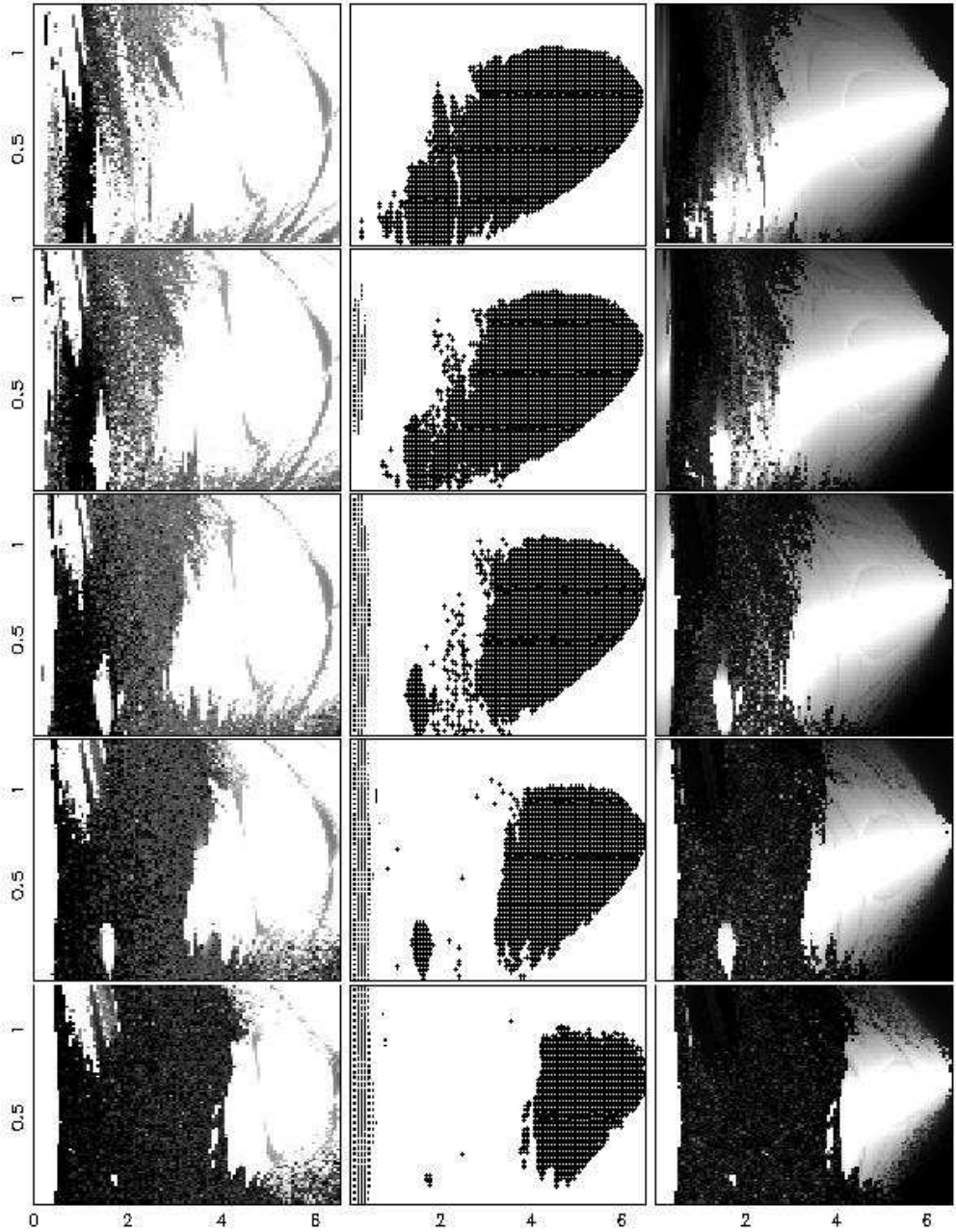}
\caption{Double-bar galaxy models with secondary bar masses increasing
       from top to bottom by a factor of 2 each. The middle panels (third from
       the top) represent the generic model. The abscissa refers to radii
       (in kpc) and the ordinate to fractions of the local rotation velocity
       (not including the bar masses). \underline{\it Left column:\/} Grayshades show the
       logarithms of Liapunov exponents (black-to-white corresponds to
       increased stability). \underline{\it Middle column:\/} Exhibits orbits supporting
       primary (crosses) and secondary (dots) bars, respectively. \underline{\it Right
       column:\/} Maps the axial ratios of orbits (black-to-white grayshades
       correspond to increase from $p \le 1$ to $p \ge 3$). See text for more 
       details.
}
\label{fig1}
\end{figure*}

The {\it generic} model is characterized with wide regions of trajectories supporting each bar. 
In particular, almost all orbits corresponding to the secondary bar are aligned with its major 
axis ({\it middle column}) and appear regular ({\it left}), most having  axial ratios 
$p \ge 3$ ({\it right}). On the
other hand, the top panel has 4 times less massive secondary bar, and the bar surface density ratio
of only 2, while most of the parameter space exhibits regular and trapped trajectories within the primary 
bar. In fact no orbits aligned with the secondary bar have been found for this model, meaning that for 
this value of the spatial and mass scales, the secondary bar is not dense enough to produce the orbits 
required to sustain it. As this bar increases its mass (2nd panel), these orbits materialize. 
Simultaneously, the chaotic region just outside the inner bar (bar-bar interface) expands decreasing 
orbital support for the primary bar by no longer displaying significant alignment with it, towards the 
lower panels. This {\it dual} behavior defines the critical mass fraction of the secondary bar, about 
$2\%-5\%$ of the primary bar. Corresponding surface density ratio is $\sim 4-8$. Below this mass (and 
surface density) the inner bar is not sustainable, and above --- the outer bar support is dramatically 
weakened. On the basis of the left and middle panels of Figure 1, we can safely rule out the top, the fourth and 
fifth panels. This leaves the second and the third panels --- the latter corresponding to the generic 
model. 

Majority of the trajectories aligned with the primary bar and virtually all those aligned with the 
secondary bar are parented by families analogous to the single-periodic $x_1$ family in time-independent 
single-barred systems, some by  higher order families. The symmetry of these orbits requires that one 
of the coordinates is maximal when the other is null (a variation of 10\% on the exact values was 
introduced to allow for the effect of the perturbing bar). We have verified this by recording the $y$-coordinate of 
maximal $x$-excursion and {\it vice versa}. Between 1/3 to 2/3 orbits (from the top to bottom panels) of 
trajectories have been found to be quasi-periodic regular ones, with exponential timescales of the order 
of a Hubble time or larger. They represent  about 50\% of trajectories in the generic model. Most  
($50\%-70\%$) of the regular trajectories are elongated in the direction of either bars, when viewed in 
the relevant  frame.   

However, a significant fraction of the trajectories supporting the bars includes trapped 
chaotic orbits. This is especially true in the case of the primary bar, where such trapped orbits 
constitute about 16\% of the supporting trajectories in the generic model. Their fraction 
peaks at 20\% when the secondary bar mass is halved, while they are quickly replaced by ``strongly 
chaotic'' orbits when the mass is doubled. Although the trapped trajectories 
have a non-zero Liapunov exponent, many of them mimic bar-supporting orbits
for a Hubble time or so. In general, trapped chaotic 
trajectories may wander intermittently between regular and chaotic phases with a distribution 
described by non-standard statistics (Zaslavsky 2002). If the initial conditions are such that a 
significant number of these trajectories are in a trapped phase, they may be of crucial importance 
to building such systems as double-barred galaxies. This issue will be elaborated elsewhere.

\section{Discussion and Conclusions}

Observations suggest that galaxies with nested bars are not an exceptional but rather a common
phenomenon. However, dynamical and evolutionary consequences of such systems are still
unclear in comparison e.g., with important effects of the large-scale stellar bars on galaxy 
evolution. The steadily increasing spatial resolution of multiwavelength observations will shortly 
provide the necessary details of stellar and gas kinematics and distribution of star formation 
in secondary bars confined to the central kpc. Here we outline the
parameter space restricted to such long-lived systems.
Based on the well-studied, simple but generic example of a double-well oscillator, 
we suggest that unless one of the bars dominates the gravitational 
potential, the trajectories 
should shift erratically between the potential wells formed by the bars, with one or both 
structures dissolving, thus contradicting the high frequency of double-barred galaxies observed 
in the local universe. 

Indeed, in a somewhat analogous situation of a bar embedded in a 
non-rotating triaxial halo, we have shown that the bar is unsustainable unless its contribution 
significantly exceeds that of the halo, in which case it is able to trap its supporting orbits 
and stabilize (El-Zant \& Shlosman 2002). In nested bar systems a similar situation can be 
constructed by invoking  separation of mass and length scales, with each bar dominating 
its own domain. In the case of spatial scales
there seems, in fact, to be observational support for this thesis. 
Laine et al. (2002) find a bimodal length distribution of bars in double-barred galaxies, with 
secondary bars confined to within 12\% of galactic radius (given by $D_{25}/2$). More
precisely, while large stellar bars are found to correlate linearly with the disk size, 
secondary bars do not exhibit this property. A straightforward explanation of 
this phenomenon is that secondary bars are confined to within the ILR of the primary bar. The ILR 
is expected to be located where the 3-dimensional nature of the 
disk cannot be ignored, i.e., at about the bulge radius for early-type (S0-Sb) disks and at about 
1~kpc for the late types, where the disk thickness becomes comparable to its radius. This 
confirms theoretical expectations that ILR serves as a dynamical separator between the bars.  

Concurrently, a {\it critical} mass necessarily exists for the secondary bar. Below its value, this
bar does not generate supporting orbital families, as it
is not dense enough to be self-gravitating and to maintain a
self-consistent orbital structure. Moreover, above this critical mass,  
as we have shown here, orbits of the primary bar become substantially affected and destabilized. 
Consequently, too massive a secondary bar, with major axis extending to near ILR
of the primary, will tend to weaken and ultimately dissolve the primary one.
The implication being that {\it it is only
within a limited range of masses and linear sizes that double-barred 
systems can develop into long-lived configurations.} This raises 
the following interesting question: 
how, in practice, is a physical system guided into the limited range 
of parameter space where it posseses the right phase space  
structure necessary for its survival?

Although our conclusions are based on purely dynamical considerations, they 
nevertheless naturally fit the context of a broader physical picture. 
How can the separation of scales be achieved in a physical configuration?
In a pure stellar system, this probably can be obtained through a 
restrictive set of initial conditions only, i.e., the system is preset to develop double bars 
(e.g., Friedli \& Martinet 1993). A more general way is to invoke the gas redistribution in
the galaxy and its accumulation within the central few hundred pc as a precondition to 
formation and dynamical decoupling of the secondary bars (e.g., Shlosman et al. 1989; Friedli 
\& Martinet 1993; Knapen et al. 1995; Heller, Shlosman \& Englmaier 2001). The secondary bars in 
this latter scenario require a gravitational runaway in the gas to initiate the decoupling. 
When the gas gravity triggers the cascade of smaller bars, the phenomena is expected to be
transient due to 
the finite gas supply and dissipation present. Stellar secondary bars in this picture are 
by-products of the runaway, through stellar capture and induced star formation in the gas.  
These two processes can regulate the parameters of the 
inner bar so as to be in a critical state of ``marginal" self-gravity
which also happens to be dynamically  long-lived. This can explain the remarkable 
frequency of double-barred systems, depite their {\it a priori} improbability.

In this Letter we have avoided varying the pattern speeds of the bars, assuming that the primary
bar  extends to near corotation and the secondary comprises 
about 2/3 of its corotation, as argued in Shlosman \& Heller (2002).
We also expect that slower rotating secondary bars, with pattern speeds closer to those
of the primary ones, will generate more orbital instability, and will, therefore, tend to 
destroy more of the regular trajectories. This effect is similar to the well known adiabatic
invariance (e.g., Arnold 1989), namely, when frequencies differ substantially, averaging over a fast 
frequency allows one to neglect, to  first order, the imposed perturbation.

Finally, it is important to emphasize that the schematic picture whereas bar-supporting  orbits 
are completely regular is an idealization. As we find, in reality many ``trapped'' chaotic 
trajectories, constrained in shape to fit the bar pattern for many rotations,  also contribute. 
Enigmatic issues related to their preponderance and importance have been known for decades 
(e.g., Goodman \& Schwarzschild 1981). Yet there is still no systematic manner of characterizing their 
evolution. We have touched on this role here in a heuristic  manner, by employing diagnostics 
such as orbital axis ratios and maximal extensions within a limited time period. The 
complex time-dependent structure of these trajectories will be examined elsewhere, by analyzing the 
time correlations of orbital segments.
 
\acknowledgments
This work was supported in part by NASA grants NAG 5-3841, NAG 5-13063, WKU-516140-02-07 and 
HST GO-08123.01-97A (provided by NASA through a grant from the STScI, which is operated by the 
AURA, under NASA contract NAS5-26555), and AST 02-06251.


 



\end{document}

%% file: epsfig.tex
\def\fileversion{v1.20a}
\def\filedate{21.6.94}
\edef\epsfigRestoreAt{\catcode`@=\number\catcode`@\relax}%
\catcode`\@=11\relax
\ifx\undefined\@makeother                
\def\@makeother#1{\catcode`#1=12\relax}  
\fi                                      
\immediate\write16{Document style option `epsfig', \fileversion\space
<\filedate> (edited by SPQR + pks)}
\newcount\EPS@Height \newcount\EPS@Width \newcount\EPS@xscale
\newcount\EPS@yscale
\def\psfigdriver#1{%
  \bgroup\edef\next{\def\noexpand\tempa{#1}}%
    \uppercase\expandafter{\next}%
    \def\LN{DVITOLN03}%
    \def\DVItoPS{DVITOPS}%
    \def\DVIPS{DVIPS}%
    \def\emTeX{EMTEX}%
    \def\OzTeX{OZTEX}%
    \def\Textures{TEXTURES}%
    \global\chardef\fig@driver=0
    \ifx\tempa\LN
        \global\chardef\fig@driver=0\fi
    \ifx\tempa\DVItoPS
        \global\chardef\fig@driver=1\fi
    \ifx\tempa\DVIPS
        \global\chardef\fig@driver=2\fi
    \ifx\tempa\emTeX
        \global\chardef\fig@driver=3\fi
    \ifx\tempa\OzTeX
        \global\chardef\fig@driver=4\fi
    \ifx\tempa\Textures
        \global\chardef\fig@driver=5\fi
  \egroup
\def\psfig@start{}%
\def\psfig@end{}%
\def\epsfig@gofer{}%
\ifcase\fig@driver
\typeout{WARNING! ****
 no specials for LN03 psfig}%
\or 
\def\psfig@start{}%
\def\psfig@end{\special{dvitops: import \@p@sfilefinal \space
\@p@swidth sp \space \@p@sheight sp \space fill}%
\if@clip \typeout{Clipping not supported}\fi
\if@angle \typeout{Rotating not supported}\fi
}%
\let\epsfig@gofer\psfig@end
\or 
\def\psfig@start{\special{ps::[begin]  \@p@swidth \space \@p@sheight \space%
        \@p@sbbllx \space \@p@sbblly \space%
        \@p@sbburx \space \@p@sbbury \space%
        startTexFig \space }%
        \if@clip
                \if@verbose
                        \typeout{(clipped to BB) }%
                \fi
                \special{ps:: doclip \space }%
        \fi
        \if@angle              
                \special {ps:: \@p@sangle \space rotate \space}
        \fi
        \special{ps: plotfile \@p@sfilefinal \space }%
        \special{ps::[end] endTexFig \space }%
}%
\def\psfig@end{}%
\def\epsfig@gofer{\if@clip
                        \if@verbose
                           \typeout{(clipped to BB)}%
                        \fi
                        \epsfclipon
                  \fi
                  \epsfsetgraph{\@p@sfilefinal}%
}%
\or 
\typeout{WARNING. You must have a .bb info file with the Bounding Box
  of the pcx file}%
\def\psfig@start{}%
\def\psfig@end{\typeout{pcx import of \@p@sfilefinal}%
\if@clip \typeout{Clipping not supported}\fi
\if@angle \typeout{Rotating not supported}\fi
\raisebox{\@p@srheight sp}{\special{em: graph \@p@sfilefinal}}}%
\def\epsfig@gofer{}%
\or 
\def\psfig@start{}%
\def\psfig@end{%
\EPS@Width\@p@swidth
\EPS@Height\@p@sheight
\divide\EPS@Width by 65781  
\divide\EPS@Height by 65781
\special{epsf=\@p@sfilefinal
\space
width=\the\EPS@Width
\space
height=\the\EPS@Height
}%
\if@clip \typeout{Clipping not supported}\fi
\if@angle \typeout{Rotating not supported}\fi
}%
\let\epsfig@gofer\psfig@end
\or 
\def\psfig@end{
         \EPS@Width=\@bbw  
         \divide\EPS@Width by 1000
         \EPS@xscale=\@p@swidth \divide \EPS@xscale by \EPS@Width
         \EPS@Height=\@bbh  
         \divide\EPS@Height by 1000
         \EPS@yscale=\@p@sheight \divide \EPS@yscale by\EPS@Height
  \ifnum\EPS@xscale>\EPS@yscale\EPS@xscale=\EPS@yscale\fi
\if@clip
   \if@verbose
      \typeout{(clipped to BB)}%
   \fi
   \epsfclipon
\fi
\special{illustration \@p@sfilefinal\space scaled \the\EPS@xscale}%
}%
\def\psfig@start{}%
\let\epsfig\psfig
\else
\typeout{WARNING. *** unknown  driver - no psfig}%
\fi
}%
\newdimen\ps@dimcent
\ifx\undefined\fbox
\newdimen\fboxrule
\newdimen\fboxsep
\newdimen\ps@tempdima
\newbox\ps@tempboxa
\long\def\fbox#1{\leavevmode\setbox\ps@tempboxa\hbox{#1}\ps@tempdima\fboxrule
    \advance\ps@tempdima \fboxsep \advance\ps@tempdima \dp\ps@tempboxa
   \hbox{\lower \ps@tempdima\hbox
  {\vbox{\hrule height \fboxrule
          \hbox{\vrule width \fboxrule \hskip\fboxsep
          \vbox{\vskip\fboxsep \box\ps@tempboxa\vskip\fboxsep}\hskip
                 \fboxsep\vrule width \fboxrule}%
                 \hrule height \fboxrule}}}}%
\fi
\ifx\@ifundefined\undefined
\long\def\@ifundefined#1#2#3{\expandafter\ifx\csname
  #1\endcsname\relax#2\else#3\fi}%
\fi
\@ifundefined{typeout}%
{\gdef\typeout#1{\immediate\write\sixt@@n{#1}}}%
{\relax}%
\@ifundefined{epsfig}{}{\typeout{EPSFIG --- already loaded} }%
\@ifundefined{epsfbox}{\input epsf}{}%
\ifx\undefined\@latexerr
        \newlinechar`\^^J
        \def\@spaces{\space\space\space\space}%
        \def\@latexerr#1#2{%
        \edef\@tempc{#2}\expandafter\errhelp\expandafter{\@tempc}%
        \typeout{Error. \space see a manual for explanation.^^J
         \space\@spaces\@spaces\@spaces Type \space H <return> \space for
         immediate help.}\errmessage{#1}}%
\fi
\def\@whattodo{You tried to include a PostScript figure which
cannot be found^^JIf you press return to carry on anyway,^^J
The failed name will be printed in place of the figure.^^J
or type X to quit}%
\def\@whattodobb{You tried to include a PostScript figure which
has no^^Jbounding box, and you supplied none.^^J
If you press return to carry on anyway,^^J
The failed name will be printed in place of the figure.^^J
or type X to quit}%
\def\@nnil{\@nil}%
\def\@empty{}%
\def\@psdonoop#1\@@#2#3{}%
\def\@psdo#1:=#2\do#3{\edef\@psdotmp{#2}\ifx\@psdotmp\@empty \else
    \expandafter\@psdoloop#2,\@nil,\@nil\@@#1{#3}\fi}%
\def\@psdoloop#1,#2,#3\@@#4#5{\def#4{#1}\ifx #4\@nnil \else
       #5\def#4{#2}\ifx #4\@nnil \else#5\@ipsdoloop #3\@@#4{#5}\fi\fi}%
\def\@ipsdoloop#1,#2\@@#3#4{\def#3{#1}\ifx #3\@nnil
       \let\@nextwhile=\@psdonoop \else
      #4\relax\let\@nextwhile=\@ipsdoloop\fi\@nextwhile#2\@@#3{#4}}%
\def\@tpsdo#1:=#2\do#3{\xdef\@psdotmp{#2}\ifx\@psdotmp\@empty \else
    \@tpsdoloop#2\@nil\@nil\@@#1{#3}\fi}%
\def\@tpsdoloop#1#2\@@#3#4{\def#3{#1}\ifx #3\@nnil
       \let\@nextwhile=\@psdonoop \else
      #4\relax\let\@nextwhile=\@tpsdoloop\fi\@nextwhile#2\@@#3{#4}}%
\long\def\epsfaux#1#2:#3\\{\ifx#1\epsfpercent
   \def\testit{#2}\ifx\testit\epsfbblit
        \@atendfalse
        \epsf@atend #3 . \\%
        \if@atend
           \if@verbose
                \typeout{epsfig: found `(atend)'; continuing search}%
           \fi
        \else
                \epsfgrab #3 . . . \\%
                \epsffileokfalse\global\no@bbfalse
                \global\epsfbbfoundtrue
        \fi
   \fi\fi}%
\def\epsf@atendlit{(atend)}
\def\epsf@atend #1 #2 #3\\{%
   \def\epsf@tmp{#1}\ifx\epsf@tmp\empty
      \epsf@atend #2 #3 .\\\else
   \ifx\epsf@tmp\epsf@atendlit\@atendtrue\fi\fi}%

\chardef\trig@letter = 11
\chardef\other = 12
 
\newif\ifdebug 
\newif\ifc@mpute 
\newif\if@atend
\c@mputetrue 
 
\let\then = \relax
\def\r@dian{pt }%
\let\r@dians = \r@dian
\let\dimensionless@nit = \r@dian
\let\dimensionless@nits = \dimensionless@nit
\def\internal@nit{sp }%
\let\internal@nits = \internal@nit
\newif\ifstillc@nverging
\def \Mess@ge #1{\ifdebug \then \message {#1} \fi}%
 
{ 
        \catcode `\@ = \trig@letter
        \gdef \nodimen {\expandafter \n@dimen \the \dimen}%
        \gdef \term #1 #2 #3%
               {\edef \t@ {\the #1}
                \edef \t@@ {\expandafter \n@dimen \the #2\r@dian}%
                \t@rm {\t@} {\t@@} {#3}%
               }%
        \gdef \t@rm #1 #2 #3%
               {{%
                \count 0 = 0
                \dimen 0 = 1 \dimensionless@nit
                \dimen 2 = #2\relax
                \Mess@ge {Calculating term #1 of \nodimen 2}%
                \loop
                \ifnum  \count 0 < #1
                \then   \advance \count 0 by 1
                        \Mess@ge {Iteration \the \count 0 \space}%
                        \Multiply \dimen 0 by {\dimen 2}%
                        \Mess@ge {After multiplication, term = \nodimen 0}%
                        \Divide \dimen 0 by {\count 0}%
                        \Mess@ge {After division, term = \nodimen 0}%
                \repeat
                \Mess@ge {Final value for term #1 of
                                \nodimen 2 \space is \nodimen 0}%
                \xdef \Term {#3 = \nodimen 0 \r@dians}%
                \aftergroup \Term
               }}%
        \catcode `\p = \other
        \catcode `\t = \other
        \gdef \n@dimen #1pt{#1} 
}%
 
\def \Divide #1by #2{\divide #1 by #2} 
 
\def \Multiply #1by #2
       {{
        \count 0 = #1\relax
        \count 2 = #2\relax
        \count 4 = 65536
        \Mess@ge {Before scaling, count 0 = \the \count 0 \space and
                        count 2 = \the \count 2}%
        \ifnum  \count 0 > 32767 
        \then   \divide \count 0 by 4
                \divide \count 4 by 4
        \else   \ifnum  \count 0 < -32767
                \then   \divide \count 0 by 4
                        \divide \count 4 by 4
                \else
                \fi
        \fi
        \ifnum  \count 2 > 32767 
        \then   \divide \count 2 by 4
                \divide \count 4 by 4
        \else   \ifnum  \count 2 < -32767
                \then   \divide \count 2 by 4
                        \divide \count 4 by 4
                \else
                \fi
        \fi
        \multiply \count 0 by \count 2
        \divide \count 0 by \count 4
        \xdef \product {#1 = \the \count 0 \internal@nits}%
        \aftergroup \product
       }}%
 
\def\r@duce{\ifdim\dimen0 > 90\r@dian \then   
                \multiply\dimen0 by -1
                \advance\dimen0 by 180\r@dian
                \r@duce
            \else \ifdim\dimen0 < -90\r@dian \then  
                \advance\dimen0 by 360\r@dian
                \r@duce
                \fi
            \fi}%
 
\def\Sine#1%
       {{%
        \dimen 0 = #1 \r@dian
        \r@duce
        \ifdim\dimen0 = -90\r@dian \then
           \dimen4 = -1\r@dian
           \c@mputefalse
        \fi
        \ifdim\dimen0 = 90\r@dian \then
           \dimen4 = 1\r@dian
           \c@mputefalse
        \fi
        \ifdim\dimen0 = 0\r@dian \then
           \dimen4 = 0\r@dian
           \c@mputefalse
        \fi
        \ifc@mpute \then
                \divide\dimen0 by 180
                \dimen0=3.141592654\dimen0
                \dimen 2 = 3.1415926535897963\r@dian 
                \divide\dimen 2 by 2 
                \Mess@ge {Sin: calculating Sin of \nodimen 0}%
                \count 0 = 1 
                \dimen 2 = 1 \r@dian 
                \dimen 4 = 0 \r@dian 
                \loop
                        \ifnum  \dimen 2 = 0 
                        \then   \stillc@nvergingfalse
                        \else   \stillc@nvergingtrue
                        \fi
                        \ifstillc@nverging 
                        \then   \term {\count 0} {\dimen 0} {\dimen 2}%
                                \advance \count 0 by 2
                                \count 2 = \count 0
                                \divide \count 2 by 2
                                \ifodd  \count 2 
                                \then   \advance \dimen 4 by \dimen 2
                                \else   \advance \dimen 4 by -\dimen 2
                                \fi
                \repeat
        \fi
                        \xdef \sine {\nodimen 4}%
       }}%
 
\def\Cosine#1{\ifx\sine\UnDefined\edef\Savesine{\relax}\else
                             \edef\Savesine{\sine}\fi
        {\dimen0=#1\r@dian\multiply\dimen0 by -1
         \advance\dimen0 by 90\r@dian
         \Sine{\nodimen 0}%
         \xdef\cosine{\sine}%
         \xdef\sine{\Savesine}}}
\def\psdraft{\def\@psdraft{0}}%
\def\psfull{\def\@psdraft{1}}%
\psfull
\newif\if@compress
\def\pscompress{\@compresstrue}
\def\psnocompress{\@compressfalse}
\@compressfalse
\newif\if@scalefirst
\def\psscalefirst{\@scalefirsttrue}%
\def\psrotatefirst{\@scalefirstfalse}%
\psrotatefirst
\newif\if@draftbox
\def\psnodraftbox{\@draftboxfalse}%
\@draftboxtrue
\newif\if@noisy
\@noisyfalse
\newif\ifno@bb
\newif\if@bbllx
\newif\if@bblly
\newif\if@bburx
\newif\if@bbury
\newif\if@height
\newif\if@width
\newif\if@rheight
\newif\if@rwidth
\newif\if@angle
\newif\if@clip
\newif\if@verbose
\newif\if@prologfile
\def\@p@@sprolog#1{\@prologfiletrue\def\@prologfileval{#1}}%
\def\@p@@sclip#1{\@cliptrue}%
\newif\ifepsfig@dos  
\def\epsfigdos{\epsfig@dostrue}%
\epsfig@dosfalse
\newif\ifuse@psfig
\def\ParseName#1{\expandafter\@Parse#1}%
\def\@Parse#1.#2:{\gdef\BaseName{#1}\gdef\FileType{#2}}%

\def\@p@@sfile#1{%
  \ifepsfig@dos
     \ParseName{#1:}%
  \else
     \gdef\BaseName{#1}\gdef\FileType{}%
  \fi
  \def\@p@sfile{NO FILE: #1}%
  \def\@p@sfilefinal{NO FILE: #1}%
  \openin1=#1
  \ifeof1\closein1\openin1=\BaseName.bb
    \ifeof1\closein1
      \if@bbllx                 
        \if@bblly\if@bburx\if@bbury
          \def\@p@sfile{#1}%
          \def\@p@sfilefinal{#1}%
        \fi\fi\fi
      \else                     
        \@latexerr{ERROR. PostScript file #1 not found}\@whattodo
        \@p@@sbbllx{100bp}%
        \@p@@sbblly{100bp}%
        \@p@@sbburx{200bp}%
        \@p@@sbbury{200bp}%
        \psdraft
      \fi
    \else                       
      \closein1%
      \edef\@p@sfile{\BaseName.bb}%
      \typeout{using BB from \@p@sfile}%
      \ifnum\fig@driver=3
        \edef\@p@sfilefinal{\BaseName.pcx}%
      \else
        \ifepsfig@dos
          \edef\@p@sfilefinal{"`gunzip -c `texfind \BaseName.{z,Z,gz}"}%
        \else
          \edef\@p@sfilefinal{"`epsfig \if@compress-c \fi#1"}%
        \fi
      \fi
    \fi
  \else\closein1                
    \edef\@p@sfile{#1}%
    \if@compress  
      \edef\@p@sfilefinal{"`epsfig -c #1"}%
    \else
      \edef\@p@sfilefinal{#1}%
    \fi
  \fi%
}

\let\@p@@sfigure\@p@@sfile
\def\@p@@sbbllx#1{%
                                            \@bbllxtrue
                \ps@dimcent=#1
                \edef\@p@sbbllx{\number\ps@dimcent}%
                \divide\ps@dimcent by65536
                \global\edef\epsfllx{\number\ps@dimcent}%
}%
\def\@p@@sbblly#1{%
                \@bbllytrue
                \ps@dimcent=#1
                \edef\@p@sbblly{\number\ps@dimcent}%
                \divide\ps@dimcent by65536
                \global\edef\epsflly{\number\ps@dimcent}%
}%
\def\@p@@sbburx#1{%
                \@bburxtrue
                \ps@dimcent=#1
                \edef\@p@sbburx{\number\ps@dimcent}%
                \divide\ps@dimcent by65536
                \global\edef\epsfurx{\number\ps@dimcent}%
}%
\def\@p@@sbbury#1{%
                \@bburytrue
                \ps@dimcent=#1
                \edef\@p@sbbury{\number\ps@dimcent}%
                \divide\ps@dimcent by65536
                \global\edef\epsfury{\number\ps@dimcent}%
}%
\def\@p@@sheight#1{%
                \@heighttrue
                \global\epsfysize=#1
                \ps@dimcent=#1
                \edef\@p@sheight{\number\ps@dimcent}%
}%
\def\@p@@swidth#1{%
                \@widthtrue
                \global\epsfxsize=#1
                \ps@dimcent=#1
                \edef\@p@swidth{\number\ps@dimcent}%
}%
\def\@p@@srheight#1{%
                \@rheighttrue\use@psfigtrue
                \ps@dimcent=#1
                \edef\@p@srheight{\number\ps@dimcent}%
}%
\def\@p@@srwidth#1{%
                \@rwidthtrue\use@psfigtrue
                \ps@dimcent=#1
                \edef\@p@srwidth{\number\ps@dimcent}%
}%
\def\@p@@sangle#1{%
                \use@psfigtrue
                \@angletrue
                \edef\@p@sangle{#1}%
}%
\def\@p@@ssilent#1{%
                \@verbosefalse
}%
\def\@p@@snoisy#1{%
                \@verbosetrue
}%
\def\@cs@name#1{\csname #1\endcsname}%
\def\@setparms#1=#2,{\@cs@name{@p@@s#1}{#2}}%
\def\ps@init@parms{%
                \@bbllxfalse \@bbllyfalse
                \@bburxfalse \@bburyfalse
                \@heightfalse \@widthfalse
                \@rheightfalse \@rwidthfalse
                \def\@p@sbbllx{}\def\@p@sbblly{}%
                \def\@p@sbburx{}\def\@p@sbbury{}%
                \def\@p@sheight{}\def\@p@swidth{}%
                \def\@p@srheight{}\def\@p@srwidth{}%
                \def\@p@sangle{0}%
                \def\@p@sfile{}%
                \use@psfigfalse
                \@prologfilefalse
                \def\@sc{}%
                \if@noisy
                        \@verbosetrue
                \else
                        \@verbosefalse
                \fi
                \@clipfalse
}%
\def\parse@ps@parms#1{%
                \@psdo\@psfiga:=#1\do
                   {\expandafter\@setparms\@psfiga,}%
\if@prologfile
\fi
}%
\def\bb@missing{%
        \if@verbose
            \typeout{psfig: searching \@p@sfile \space  for bounding box}%
        \fi
        \epsfgetbb{\@p@sfile}%
        \ifepsfbbfound
            \ps@dimcent=\epsfllx bp\edef\@p@sbbllx{\number\ps@dimcent}%
            \ps@dimcent=\epsflly bp\edef\@p@sbblly{\number\ps@dimcent}%
            \ps@dimcent=\epsfurx bp\edef\@p@sbburx{\number\ps@dimcent}%
            \ps@dimcent=\epsfury bp\edef\@p@sbbury{\number\ps@dimcent}%
        \else
            \epsfbbfoundfalse
        \fi
}
\newdimen\p@intvaluex
\newdimen\p@intvaluey
\def\rotate@#1#2{{\dimen0=#1 sp\dimen1=#2 sp
                  \global\p@intvaluex=\cosine\dimen0
                  \dimen3=\sine\dimen1
                  \global\advance\p@intvaluex by -\dimen3
                  \global\p@intvaluey=\sine\dimen0
                  \dimen3=\cosine\dimen1
                  \global\advance\p@intvaluey by \dimen3
                  }}%
\def\compute@bb{%
                \epsfbbfoundfalse
                \if@bbllx\epsfbbfoundtrue\fi
                \if@bblly\epsfbbfoundtrue\fi
                \if@bburx\epsfbbfoundtrue\fi
                \if@bbury\epsfbbfoundtrue\fi
                \ifepsfbbfound\else\bb@missing\fi
                \ifepsfbbfound\else
                \@latexerr{ERROR. cannot locate BoundingBox}\@whattodobb
                        \@p@@sbbllx{100bp}%
                        \@p@@sbblly{100bp}%
                        \@p@@sbburx{200bp}%
                        \@p@@sbbury{200bp}%
                        \no@bbtrue
                        \psdraft
                \fi
                \count203=\@p@sbburx
                \count204=\@p@sbbury
                \advance\count203 by -\@p@sbbllx
                \advance\count204 by -\@p@sbblly
                \edef\ps@bbw{\number\count203}%
                \edef\ps@bbh{\number\count204}%
                 \edef\@bbw{\number\count203}%
                \edef\@bbh{\number\count204}%
               \if@angle
                        \Sine{\@p@sangle}\Cosine{\@p@sangle}%
 
{\ps@dimcent=\maxdimen\xdef\r@p@sbbllx{\number\ps@dimcent}%
 
\xdef\r@p@sbblly{\number\ps@dimcent}%
 
\xdef\r@p@sbburx{-\number\ps@dimcent}%
 
\xdef\r@p@sbbury{-\number\ps@dimcent}}%
                        \def\minmaxtest{%
                           \ifnum\number\p@intvaluex<\r@p@sbbllx
                              \xdef\r@p@sbbllx{\number\p@intvaluex}\fi
                           \ifnum\number\p@intvaluex>\r@p@sbburx
                              \xdef\r@p@sbburx{\number\p@intvaluex}\fi
                           \ifnum\number\p@intvaluey<\r@p@sbblly
                              \xdef\r@p@sbblly{\number\p@intvaluey}\fi
                           \ifnum\number\p@intvaluey>\r@p@sbbury
                              \xdef\r@p@sbbury{\number\p@intvaluey}\fi
                           }%
                        \rotate@{\@p@sbbllx}{\@p@sbblly}%
                        \minmaxtest
                        \rotate@{\@p@sbbllx}{\@p@sbbury}%
                        \minmaxtest
                        \rotate@{\@p@sbburx}{\@p@sbblly}%
                        \minmaxtest
                        \rotate@{\@p@sbburx}{\@p@sbbury}%
                        \minmaxtest
 
\edef\@p@sbbllx{\r@p@sbbllx}\edef\@p@sbblly{\r@p@sbblly}%
 
\edef\@p@sbburx{\r@p@sbburx}\edef\@p@sbbury{\r@p@sbbury}%
                \fi
                \count203=\@p@sbburx
                \count204=\@p@sbbury
                \advance\count203 by -\@p@sbbllx
                \advance\count204 by -\@p@sbblly
                \edef\@bbw{\number\count203}%
                \edef\@bbh{\number\count204}%
}%
\def\in@hundreds#1#2#3{\count240=#2 \count241=#3
                     \count100=\count240        
                     \divide\count100 by \count241
                     \count101=\count100
                     \multiply\count101 by \count241
                     \advance\count240 by -\count101
                     \multiply\count240 by 10
                     \count101=\count240        
                     \divide\count101 by \count241
                     \count102=\count101
                     \multiply\count102 by \count241
                     \advance\count240 by -\count102
                     \multiply\count240 by 10
                     \count102=\count240        
                     \divide\count102 by \count241
                     \count200=#1\count205=0
                     \count201=\count200
                        \multiply\count201 by \count100
                        \advance\count205 by \count201
                     \count201=\count200
                        \divide\count201 by 10
                        \multiply\count201 by \count101
                        \advance\count205 by \count201
                     \count201=\count200
                        \divide\count201 by 100
                        \multiply\count201 by \count102
                        \advance\count205 by \count201
                     \edef\@result{\number\count205}%
}%
\def\compute@wfromh{%
                \in@hundreds{\@p@sheight}{\@bbw}{\@bbh}%
                \edef\@p@swidth{\@result}%
}%
\def\compute@hfromw{%
                \in@hundreds{\@p@swidth}{\@bbh}{\@bbw}%
                \edef\@p@sheight{\@result}%
}%
\def\compute@handw{%
                \if@height
                        \if@width
                        \else
                                \compute@wfromh
                        \fi
                \else
                        \if@width
                                \compute@hfromw
                        \else
                                \edef\@p@sheight{\@bbh}%
                                \edef\@p@swidth{\@bbw}%
                        \fi
                \fi
}%
\def\compute@resv{%
                \if@rheight \else \edef\@p@srheight{\@p@sheight} \fi
                \if@rwidth \else \edef\@p@srwidth{\@p@swidth} \fi
}%
\def\compute@sizes{%
        \if@scalefirst\if@angle
        \if@width
           \in@hundreds{\@p@swidth}{\@bbw}{\ps@bbw}%
           \edef\@p@swidth{\@result}%
        \fi
        \if@height
           \in@hundreds{\@p@sheight}{\@bbh}{\ps@bbh}%
           \edef\@p@sheight{\@result}%
        \fi
        \fi\fi
        \compute@handw
        \compute@resv
}
\long\def\graphic@verb#1{\def\next{#1}%
  {\expandafter\graphic@strip\meaning\next}}
\def\graphic@strip#1>{}
\def\graphic@zapspace#1{%
  #1\ifx\graphic@zapspace#1\graphic@zapspace%
  \else\expandafter\graphic@zapspace%
  \fi}
\def\psfig#1{%
\edef\@tempa{\graphic@zapspace#1{}}%
\ifvmode\leavevmode\fi\vbox {%
        \ps@init@parms
        \parse@ps@parms{\@tempa}%
        \ifnum\@psdraft=1
                \typeout{[\@p@sfilefinal]}%
                \if@verbose
                        \typeout{epsfig: using PSFIG macros}%
                \fi
                \psfig@method
        \else
                \epsfig@draft
        \fi
}
}%
\def\graphic@zapspace#1{%
  #1\ifx\graphic@zapspace#1\graphic@zapspace%
  \else\expandafter\graphic@zapspace%
  \fi}
\def\epsfig#1{%
\edef\@tempa{\graphic@zapspace#1{}}%
\ifvmode\leavevmode\fi\vbox {%
        \ps@init@parms
        \parse@ps@parms{\@tempa}%
        \ifnum\@psdraft=1
          \if@angle\use@psfigtrue\fi
          {\ifnum\fig@driver=1\global\use@psfigtrue\fi}%
          {\ifnum\fig@driver=3\global\use@psfigtrue\fi}%
          {\ifnum\fig@driver=4\global\use@psfigtrue\fi}%
          {\ifnum\fig@driver=5\global\use@psfigtrue\fi}%
                \ifuse@psfig
                        \if@verbose
                                \typeout{epsfig: using PSFIG macros}%
                        \fi
                        \psfig@method
                \else
                        \if@verbose
                                \typeout{epsfig: using EPSF macros}%
                        \fi
                        \epsf@method
                \fi
        \else
                \epsfig@draft
        \fi
}%
}%

\def\epsf@method{%
        \epsfbbfoundfalse
        \if@bbllx\epsfbbfoundtrue\fi
        \if@bblly\epsfbbfoundtrue\fi
        \if@bburx\epsfbbfoundtrue\fi
        \if@bbury\epsfbbfoundtrue\fi
        \ifepsfbbfound\else\epsfgetbb{\@p@sfile}\fi
        \ifepsfbbfound
           \typeout{<\@p@sfilefinal>}%
           \epsfig@gofer
        \else
          \@latexerr{ERROR - Cannot locate BoundingBox}\@whattodobb
          \@p@@sbbllx{100bp}%
          \@p@@sbblly{100bp}%
          \@p@@sbburx{200bp}%
          \@p@@sbbury{200bp}%
                \count203=\@p@sbburx
                \count204=\@p@sbbury
                \advance\count203 by -\@p@sbbllx
                \advance\count204 by -\@p@sbblly
                \edef\@bbw{\number\count203}%
                \edef\@bbh{\number\count204}%
          \compute@sizes
          \epsfig@@draft
       \fi
}%
\def\psfig@method{%
        \compute@bb
        \ifepsfbbfound
          \compute@sizes
          \psfig@start
          \vbox to \@p@srheight sp{\hbox to \@p@srwidth 
            sp{\hss}\vss\psfig@end}%
        \else
           \epsfig@draft
        \fi
}%
\def\epsfig@draft{\compute@bb\compute@sizes\epsfig@@draft}%
\def\epsfig@@draft{%
\typeout{<(draft only) \@p@sfilefinal>}%
\if@draftbox
        \hbox{{\fboxsep0pt\fbox{\vbox to \@p@srheight sp{%
        \vss\hbox to \@p@srwidth sp{ \hss 
           \expandafter\Literally\@p@sfilefinal\@nil
                          \hss }\vss
        }}}}%
\else
        \vbox to \@p@srheight sp{%
        \vss\hbox to \@p@srwidth sp{\hss}\vss}%
\fi
}%
\def\Literally#1\@nil{{\tt\graphic@verb{#1}}}
\psfigdriver{dvips}%
\epsfigRestoreAt